\documentclass[12pt]{article}
\usepackage{graphics}
\usepackage{bm}
\usepackage{graphicx}
\usepackage{amsmath}
\usepackage{amssymb}
\usepackage{amstext}
\usepackage{amscd}
\usepackage{amsfonts}
\usepackage{textcomp}
\usepackage{appendix}
\usepackage{color}
\DeclareMathAlphabet{\EuFrak}{U}{euf}{m}{n}
\DeclareMathAlphabet{\EuScript}{U}{eus}{m}{n}

\newcommand{\nd}{\noindent}

\title{{\bf Tsallis' Quantum q-Fields}}
\author{{A. Plastino$^{1,3,4}$, M.C.Rocca$^{1,2,3}$}, \\
\small{$^1$ Departamento de F\'{\i}sica,
Universidad Nacional de La Plata,}\\
\small{$^2$ Departamento de Matem\'{a}tica,
Universidad Nacional de La Plata,}\\
\small{$^3$ Consejo Nacional de Investigaciones Cient\'{\i}ficas
y Tecnol\'{o}gicas}\\
\small{(IFLP-CCT-CONICET)-C. C. 727, 1900 La Plata -
Argentina}\\\small{$^4$  SThAR - EPFL, Lausanne, Switzerland}}

\date{\today}

\begin{document}

\maketitle

\begin{abstract}

\nd We generalize several well known  quantum equations to  a Tsallis' q-scenario, and provide a quantum 
version of some classical fields associated to them in recent literature. We refer to the
 q-Schr\"{o}dinger, q-Klein-Gordon, q-Dirac, and q-Proca  equations advanced in, respectively,
[Phys. Rev. Lett. {\bf 106}, 140601 (2011), EPL {\bf 118}, 61004 (2017) and references therein].
 Also, we introduce here equations corresponding to
q-Yang-Mills fields, both in the Abelian and not-Abelian instances.

\nd We show how to define the q-Quantum Field Theories corresponding to the above equations, introduce the pertinent actions, and obtain motion equations via the minimum action principle.

\nd These q-fields are meaningful at very high energies (TeVs) for
$q=1.15$, high ones  (GeVs) for $q=1.001$, and low energies
(MeVs)for  $q=1.000001$ [Nucl. Phys. A {\bf 955}  (2016) 16 and references therein]. (See the Alice experiment of LHC).

\nd Surprisingly enough, these q-fields are simultaneously q-exponential functions of the usual linear fields' logarithms.

\nd {\bf Keywords:}Non-linear Klein-Gordon,
non-linear Schr\"{o}dinger and non-linear q-Dirac fields;
Non-linear q-Yang-Mills and non-linear q-Proca fields;
Classical Field Theory; Quantum Field Theory.\\
\nd {\bf PACS:} 11.10.Ef; 11.10.Lm; 02.30.Jm

\end{abstract}

\newpage

\renewcommand{\theequation}{\arabic{section}.\arabic{equation}}

\setcounter{equation}{0}

\section{Introduction}

\nd  Classical fields theories (CFT) associated to  Tsallis' q-scenarios have
received much attention recently
\cite{tp1,ts5,tz1,tz3,tz2,tp0}. Associated q-Quantum Field Theories
(q-QFTs) have also been discussed \cite{ts5,tz1}.

\nd  These CFTs cannot be directly quantified because of non linearity,
 that makes not applicable the superposition principle, and then, impossible to introduce creation/annihilation operators for the q-fields. We will here  remedy such formidable quantification obstacle by recourse to an indirect approach.

\nd  Thus, in this paper we both extend to the quantum realm  and generalize several aspects of the above mentioned works.
 We construct the CFTs corresponding to the
 q-Schr\"{o}dinger, q-Klein-Gordon, and  q-Dirac equations introduced
in \cite{tp1,ts5,tz1,tz3}. We do the same for the q-Proca y
q-Yang-Mills (Abelian) defined in \cite{tz2}. Also, and for the
first time ever, we deal with the equation and q-QFT corresponding
to a non Abelian q-Yang-Mills      field. It has been shown in
\cite{ts1,ts2} that q-fields emerge at 1) very high energies (TeV)
for $q=1.15$, 2) High (GeV) for $q=1.001$, and 3) low (MeV) for
$q=1.000001$. LHC-Alice experiments show that Tsallis q-effects
manifest themselves  \cite{ts3} at TeV energies.

\nd We will see that all  q-QFTs employed here transform into the well known
associated  QFTs for
$q\rightarrow 1$, entailing going down from extremely high energies to lower ones.

\nd Our new quantum field theories correspond to non linear
equations. Thus, gauge and Lorentz invariance are broken. These
invariance reappear in the limit  $q\rightarrow 1$. A nice property
of our new equation $\partial_\mu A^\mu=0$,
 valid for Abelian  Yang-Mills and Proca fields are also valid for
q-Abelian Yang-Mills and q-Proca fields.\vskip 2mm

 \nd M.A. Rego-Monteiro et al. [6] have tackled in recent years the possible need of two coupled fields, instead of only one, to properly handle classical nonlinear equations. The quantification of these two coupled fields is discussed in the paper’s references [2] and [3].

 \vskip 2mm 
\nd Motivations for nonlinear quantum evolution equations can be divided
up into two types,   namely,  (A) as basic equations governing  phenomena
at the frontiers of quantum mechanics, mainly
at the boundary between quantum
 and gravitational physics (see, \cite{37,38}
and references therein). The other possibility is (B) regard nonlinear-Schr\"{o}dinger-like equations (NLSE)
as effective, single particle mean field
descriptions of involved  quantum many-body systems. A paradigmatic illustration  is that of \cite{39}.

\setcounter{equation}{0}

\section{A non-linear q-Schr\"{o}dinger Field}

\color{blue} We proceed now to effect a transformation that requires some previous considerations. 
Consider two different formalisms {\cal A} and {\cal B} that can be connected by an appropriate mathematical transformation. Assume that we know how to solve the relevant equations for  {\cal A} . A legitimate question, that one may ask is why to bother at all which formalism  {\cal B}  that could be mathematically more involved than  {\cal A} . The answer is the following.
Even though  {\cal A}  and  {\cal B}  are mathematically connected, it is possible that,  in some scenarios, the variables in {\cal B}   provide a more appropriate description on some natural phenomenon. There is some experimental evidence that such is the case with Tsallis-inspired non linear wave equation (ALICE in CERN). Empirically, they find q-exponentials, that are solutions to the q-equations of motion.  
suggesting that Nature uses the non standard scenario.
Another example refers to the Schroedinger equation (SE) with variable mass, that has multiple applications. Here there exists a transformation connecting the SE with constant mass with the SE with variable mass.  Why bother with such transformation? Answer: in many problems in solid state physics, nuclear physics, etc., the relevant physics is described by the SE with variable mass. The transformation that we are advancing here reads:

\normalcolor

\begin{equation}
\label{eq2.1}
\psi_q(\vec {x},t)=[1+(1-q)\ln\psi(\vec{x},t)]_+^{\frac {1} {1-q}},
\end{equation}
where $\psi$ is the usual quantum Schr\"{o}dinger Field  operator, the subscript $+$ indicates the so-called Tsallis cut-off.

\color{blue}\nd  At a quantum level, which is the case that we are interested in here, the cut-off has no relevance since $\psi$ is an operator and the information is contained in the pertinent  operators of creation and annihilation. At this level we have:
\[\psi_q=[I+(1-q)\ln\psi]_+^{\frac {1} {1-q}}=
e^{\frac {1} {1-q}\ln\{[I+(1-q)\ln\psi]\}}=\]
\begin{equation}
\label{c}
\sum\limits_{n=0}^{\infty}a_n\frac {\phi^n} {n!}=I+\phi+\frac {(q-1)} {2}\phi^2+....
\end{equation}
where $\psi=I+\phi$. There are no cuts or branch points then. No information is lost if one considers the whole series.\normalcolor

\vskip 3mm \nd Consider  now the classical instance in which  $\psi$ is just a plane wave

\begin{equation}
\label{eq2.2}
\psi(\vec {x},t)=e^{\frac {i} {\hbar}(\vec{p}\cdot\vec{x}-Et)}.
\end{equation}
Replacing this into  (\ref{eq2.1}) we find
\begin{equation}
\label{eq2.3}
\psi_q(\vec{x},t)=[1+(1-q)\frac {i} {\hbar}
(\vec{p}\cdot\vec{x}-Et)]_+^{\frac {1} {1-q}}.
\end{equation}
This is just the q-wave of  Nobre et al.
 \cite{tp1} used there to obtain  q-Schr\"{o}dinger,
q-Klein-Gordon, and q-Dirac equations. Thus, the q-wave is a particular case of the quantum field defined
by (\ref{eq2.1}). This allows for immediate generalization to the quantum realm of  the classical treatment of fields given in  \cite{tp1}. Accordingly, we can obtain quantum  q-fields starting from the usual $q=1$-usual quantum fields.
We can also express  $\psi$ in terms of $\psi_q$   as
\begin{equation}
\label{eq2.4}
\psi=e^{\frac {\psi_q^{(1-q)}-1} {1-q}}
\end{equation}
 Schr\"{o}dinger's field's action  ${\cal S}$ is well known to be
\begin{equation}
\label{eq2.5}
{\cal S}=\int\left(i\hbar \psi^{\dagger}\partial_t\psi-
\frac {\hbar^2} {2m}\nabla\psi^{\dagger}\nabla\psi\right)
dt\;d^3x.
\end{equation}
From it one deduces the motion equation
\begin{equation}
\label{eq2.6}
i\hbar\partial_t\psi+
\frac {\hbar^2} {2m}\bigtriangleup\psi=0,
\end{equation}
whose solution is
\begin{equation}
\label{eq2.7}
\psi(\vec{x},t)=\frac {1} {(2\pi\hbar)^{\frac {3} {2}}}
\int a(\vec{p})
e^{\frac {i} {\hbar}(\vec{p}\cdot\vec{x}-Et)}d^3p.
\end{equation}
The action corresponding to the field  $\psi_q$ is
\begin{equation}
\label{eq2.8}
{\cal S}_q=\int e^{\frac {\psi_q^{(1-q)}-1} {1-q}}
e^{\frac {\psi_q^{\dagger(1-q)}-1} {1-q}}
\psi_q^{-q}
\left(i\hbar\partial_t\psi-
\frac {\hbar^2} {2m}
\psi_q^{\dagger-q}
\nabla\psi_q^{\dagger}\nabla\psi_q\right)
dt\;d^3x,
\end{equation}
constructed keeping in mind that the field
$\psi_q$ satisfies
\begin{equation}
\label{eq2.9}
i\hbar\partial_t\psi_q+\frac {\hbar^2} {2m}[
\bigtriangleup\psi_q+(\nabla\psi_q)^2
(\psi_q^{-q}-q\psi_q^{-1})]=0.
\end{equation}
Note that the q-exponential wave  (\ref{eq2.5}) is, by
construction, solution to the equation  (\ref{eq2.9}).
For $q\rightarrow 1$ this last equation becomes the usual Schr\"{o}dinger equation.
Same for the action given by  (\ref{eq2.8}).
One is then in a position to assert that such an action is the q-generalization
of the usual one and that  (\ref{eq2.9})
is the q-generalization of the ordinary Schr\"{o}dinger equation.

\vskip 2mm \nd
Additionally, see that since in (\ref{eq2.1}) the field  $\psi$
 is a quantum field, this implies that $\psi_q$ is of such a nature too.
Of course, for  $q\rightarrow 1$, $\psi_q$ becomes $\psi$.
Physically, if the energy gets down, the q-field transforms itself
into the usual one (remember our assertions above on the connection
between  q-fields and the energy scale based on Alice work on LHC
\cite{ts1,ts2}). Given that we speak here of a non linear  QFT,
direct field quantification by appeal to creation-destruction
operators is not feasible, since the superposition principle is no
longer valid.  The reasoning applies to the propagator notion as
well. Thus, as we did here, an indirect route is necessary to
quantify a classical field.

\setcounter{equation}{0}

\section{A non-linear q-Klein-Gordon (KG) Field}

In the same vein as above, we define a quantum q-KG field
$\phi_q(x_\mu)$  Field  en in terms of the ordinary KG one
$\phi(x_\mu)$ as
\begin{equation}
\label{eq3.1}
\phi_q(x_\mu)=[1+(1-q)\ln\phi(x_\mu)]_+^{\frac {1} {1-q}}.
\end{equation}
In the classical instance, if we have
\begin{equation}
\label{eq3.2}
\phi(x_\mu)=e^{i(\vec{k}\cdot\vec{x}-\omega t)},
\end{equation}
we re-obtain the q-wave used by Nobre et al. in \cite{tp1}:
\begin{equation}
\label{eq3.3}
\phi_q(x_\mu)=[1+(1-q)i
(\vec{k}\cdot\vec{x}-\omega t)]_+^{\frac {1} {1-q}}.
\end{equation}
 $\phi$ can be given in terms of  $\phi_q$ as
\begin{equation}
\label{eq3.4}
\phi=e^{\frac {\phi_q^{(1-q)}-1} {1-q}}.
\end{equation}
From (\ref{eq3.1}) we see that
 $\phi_q$ is not Lorentz invariant (LI). We saw above that it manifests itself at very high energy. If the energy becomes smaller, and this happens for  $q\rightarrow 1$,  $\phi_q$ becomes
$\phi$ and LI is restored. The usual KG action is

\begin{equation}
\label{eq3.5}
{\cal S}=\int\left[
\partial_{\mu}\phi(x_{\mu})
\partial^{\mu}\phi^{\dagger}(x_{\mu})-
m^2
\phi(x_{\mu})\phi^{\dagger}(x_{\mu})\right]d^4x_\mu,
\end{equation}
from which one deduces
\begin{equation}
\label{eq3.6}
(\Box+m^2)\phi=0,
\end{equation}
whose solution is
\begin{equation}
\label{eq3.7}
\phi(x_\mu)=\frac {1} {(2\pi)^{\frac {3} {2}}}
\int
\frac {a(\vec{k})} {\sqrt{2\omega}}
e^{i(\vec{k}\cdot\vec{x}-\omega t)}+
\frac {a^{\dagger}(\vec{k})} {\sqrt{2\omega}}
e^{-i(\vec{k}\cdot\vec{x}-\omega t)}d^3k,
\end{equation}
this being the field  $\phi$ in (\ref{eq3.1}).
For  $\phi_q$ one has
\begin{equation}
\label{eq3.8}
{\cal S}_q=\int e^{\frac {\phi_q^{(1-q)}-1} {1-q}}
e^{\frac {\phi_q^{\dagger(1-q)}-1} {1-q}}
\left(\phi_q^{-q}\phi_q^{\dagger -q}
\partial_\mu\phi_q\partial_\mu\phi_q^{\dagger}-m^2
\right)d^4x_\mu,
\end{equation}
leading to an equation of motion whose solution is $\phi_q$, that is
\begin{equation}
\label{eq3.9}
\Box\phi_q+\partial_\mu\phi_q\partial^{\mu}\phi_q
(\phi_q^{-q}-q\phi_q^{-1})+m^2\phi_q^q=0.
\end{equation}
For $q\rightarrow 1$, (\ref{eq3.8})
becomes (\ref{eq3.5}) while (\ref{eq3.9}) goes over to  (\ref{eq3.6}).

\setcounter{equation}{0}

\section{A non-linear q-Dirac's Field}

Dirac's action is known to be:
\begin{equation}
\label{eq4.1}
{\cal S}=\int i\overline{\psi}\slash\hspace{-2.5mm}{\partial}\psi
-m\overline{\psi}\psi d^4x,
\end{equation}
or
\begin{equation}
\label{eq4.2}
{\cal S}=\int i\psi^\dagger\gamma^0\gamma^\mu{\partial}_\mu\psi
-m\psi^\dagger\gamma^0\psi d^4x.
\end{equation}
In terms of Dirac's spinor  $\psi$ this action is
\begin{equation}
\label{eq4.3}
{\cal S}=\int i\psi_a^\dagger(\gamma^0\gamma^\mu)_{ab}{\partial}_\mu\psi_b
-m\psi_a^\dagger\gamma_{ab}^0\psi_b d^4x.
\end{equation}
We deduce now that the spinor's components obey the equations of
motion
\begin{equation}
\label{eq4.4}
i\gamma_{ab}^\mu\partial_\mu\psi_b-m\psi_a=0,
\end{equation}
\begin{equation}
\label{eq4.5}
(\Box+m^2)\psi_a=0,
\end{equation}
that, of course, are
Dirac's and Klein-Gordon's equation, respectively. We define now a very high energy field  $\psi_{qa}$ as

\begin{equation}
\label{eq4.6}
\psi_{qa}=[1+(1-q)\ln\psi_a]_+^{\frac {1} {1-q}},
\end{equation}
not Lorentz invariant. $\psi_{qa}$ is not a component of the Diracc-spinor. Let us now cast  $\psi_a$ in
terms of
$\psi_{qa}$:
\begin{equation}
\label{eq4.7}
\psi_a=e^{\frac {\psi_{qa}^{(1-q)}-1} {1-q}}.
\end{equation}
The $\psi_{qa}$-associated action is
\[{\cal S}_q=\sum\limits_{ab}\int i
e^{\frac {\psi_{qa}^{\dagger(1-q)}-1} {1-q}}
(\gamma^0\gamma^\mu)_{ab}
\psi_{qb}^{\dagger(-q)}
{\partial}_\mu\psi_{qb}
e^{\frac {\psi_{qb}^{(1-q)}-1} {1-q}}-\]
\begin{equation}
\label{eq4.8}
m e^{\frac {\psi_{qa}^{\dagger(1-q)}-1} {1-q}}
\gamma_{ab}^0
e^{\frac {\psi_{qb}^{(1-q)}-1} {1-q}}d^4x_\mu.
\end{equation}
Given the lack of Lorentz invariance, Einstein's convention on
repeated indexes can not be used. T This action becomes that of
(\ref{eq4.3}) for  $q\rightarrow 1$. From  (\ref{eq4.8}) one
deduces the motion equations for $\psi_{q}$ as
\begin{equation}
\label{eq4.9}
i\gamma_{ab}^\mu
\psi_{qb}^{-q}
\partial_\mu\psi_{qb}
e^{\frac {\psi_{qb}^{(1-q)}-1} {1-q}}
-m e^{\frac {\psi_{qa}^{(1-q)}-1} {1-q}}=0
\end{equation}
\begin{equation}
\label{eq4.10}
\Box\psi_{qa}+\partial_\mu\psi_{qa}\partial^{\mu}\psi_{qa}
(\psi_{qa}^{-q}-q\psi_{qa}^{-1})+m^2\psi_{qa}^q=0,
\end{equation}
that become  (\ref{eq4.4}) - (\ref{eq4.5}) when
 $q\rightarrow 1$. Energetic consideration in this limit made above in the KG case also hold here.

\setcounter{equation}{0}

\section{Advancing a  non-linear Abelian Yang-Mills' q-Field}

It is well known that the action for an Abelian Yang-Mills field reads

\begin{equation}
\label{eq5.1}
{\cal S}=-\frac {1} {4}\int{\cal F}^{\mu\nu}
{\cal F}_{\mu\nu}   d^4x,
\end{equation}
where
\begin{equation}
\label{eq5.2}
{\cal F}_{\mu\nu}=\partial_\mu A_\nu-\partial_\nu A_\mu,
\end{equation}
and the associated motion equation is
\begin{equation}
\label{eq5.3}
\partial_\mu{\cal F}^{\mu\nu}=0,
\end{equation}
which can be recast as two equations
\begin{equation}
\label{eq5.4}
\Box A_\mu=0\;\;\;\;\;\partial_\mu A^\mu=0.
\end{equation}
Our present  q-extension begins by defining
\begin{equation}
\label{eq5.5}
A_{q\mu}=[1+(1-q)\ln A_\mu]_+^{\frac {1} {1-q}},
\end{equation}
breaking Lorentz invariance (LI) once again. Conversely, we can write
\begin{equation}
\label{eq5.6}
A_{\mu}=e^{\frac {A_{q\mu}^{(1-q)}-1} {1-q}},
\end{equation}
leading to
\begin{equation}
\label{eq5.7}
\partial^\mu A_{\mu}=e^{\frac {A_{q\mu}^{(1-q)}-1} {1-q}}
A_{q\mu}^{-q}\partial^{\mu}A_{q\mu}=0,
\end{equation}
and then
\begin{equation}
\label{eq5.8}
\partial^{\mu}A_{q\mu}=0,
\end{equation}
so that the field $A_{q\mu}$ fulfills  Lorentz' gauge,
a surprising result given the above LI-breaking.     Our associated q-action
 $A_{q\mu}$ is
\[{\cal S}_q=-\frac {1} {4}\sum\limits_{\mu,\nu,\rho,\eta}
g^{\mu\rho}g^{\nu\eta}\int
\left[e^{\frac {A_{q\eta}^{(1-q)}-1} {1-q}}
A_{q\eta}^{-q}\partial_\rho A_{q\eta}-
e^{\frac {A_{q\rho}^{(1-q)}-1} {1-q}}
A_{q\rho}^{-q}\partial_\eta A_{q\rho}\right]\otimes\]
\begin{equation}
\label{eq5.9}
\left[e^{\frac {A_{q\nu}^{(1-q)}-1} {1-q}}
A_{q\nu}^{-q}\partial_\mu A_{q\nu}-
e^{\frac {A_{q\mu}^{(1-q)}-1} {1-q}}
A_{q\mu}^{-q}\partial_\nu A_{q\mu}\right]d^4x,
\end{equation}
leading to the motion equation
\begin{equation}
\label{eq5.10}
\Box A_{q\mu}+\partial_\nu A_{q\mu}\partial^{\nu}A_{q\mu}
(A_{q\mu}^{-q}-q A_{q\mu}^{-1})=0,
\end{equation}
obeyed by  $A_{q\mu}$.  It is clear that for
 $q\rightarrow 1$ our new theory becomes the customary Abelian Yang-Mills one.

\setcounter{equation}{0}

\section{Introducing our  non-linear - non-Abelian Yang-Mills' q-Field}

Yang-Mills' theory is a gauge one, constructed from a Lie algebra, that attempts to  describe the behavior of elementary particles via non-Abelian Lie groups. This lies at the core of i) the unification of weak and  electromagnetic forces, as well as ii) quantum chromodynamics. It constitutes the foundation  of
our understanding of the standard model. The corresponding action is

\begin{equation}
\label{eq6.1}
{\cal S}=-\frac {1} {2g^2}\int tr(\boldsymbol{{\cal F}^{\mu\nu}}
\boldsymbol{{\cal F}_{\mu\nu}})   d^4x,
\end{equation}
where
\begin{equation}
\label{eq6.2}
\boldsymbol{{\cal F}^{\mu\nu}}=
{\cal F}^{\mu\nu}_{C} T^C.
\end{equation}
Here the matrices  $T^C$ correspond to a non abelian, semi simple Lie group. One has
\begin{equation}
\label{eq6.3}
[T_A,T_B]=f_{AB}^C T^C,
\end{equation}
\begin{equation}
\label{eq6.4}
tr(T_A T_B)=\frac {\delta_{AB}} {2},
\end{equation}
where $\boldsymbol{{\cal F}^{\mu\nu}}$ is
\begin{equation}
\label{eq6.5}
\boldsymbol{{\cal F}^{\mu\nu}}=
\partial_\mu\boldsymbol{A_\nu}-
\partial_\nu\boldsymbol{A_\mu}-ig
[\boldsymbol{A_\mu},\boldsymbol{A_\nu}],
\end{equation}
with

\begin{equation}
\label{eq6.6}
\boldsymbol{A_\mu}=A_\mu^CT_C
\end{equation}
and

\begin{equation}
\label{eq6.7}
{\cal F}_{\mu\nu}^C=\partial_\mu A_\nu^C-\partial_\nu A_\mu^C+
gf_{AB}^CA_\mu^AA_\nu^B.
\end{equation}
Because of the relation

\begin{equation}
\label{eq6.8}
tr(\boldsymbol{{\cal F}^{\mu\nu}}\boldsymbol{{\cal F}_{\mu\nu}})=
\frac {1} {2} {\cal F}^{\mu\nu C}{\cal F}_{\mu\nu C},
\end{equation}
the action becomes

\begin{equation}
\label{eq6.9}
{\cal S}=-\frac {1} {4g^2}\int {\cal F}^{\mu\nu C}
{\cal F}_{\mu\nu C}   d^4x,
\end{equation}
leading to the motion equation
\begin{equation}
\label{eq6.10}
\partial_\rho{\cal F}^{\rho\sigma D}-g
{\cal F}^{\rho\sigma C}f_{AD}^CA_{\rho}^A=0.
\end{equation}
Define now our q-extension

\begin{equation}
\label{eq6.11}
A_{q\mu}^C=[1+(1-q)\ln A_\mu^C]_+^{\frac {1} {1-q}},
\end{equation}
again breaking both  Lorentz' and gauge's invariance for  para
$q>1$. From  (\ref{eq6.11}) we obtain
\begin{equation}
\label{eq6.12}
A_{\mu}^C=e^{\frac {A_{q\mu}^C-1} {1-q}},
\end{equation}
and the action associated to the field (\ref{eq6.11})
is

\begin{equation}
\label{eq6.13}
{\cal S}=-\frac {1} {4g^2}\sum\limits_{\mu,\nu,C}g^{\mu\mu}g^{\nu\nu}
\int {\cal F}_{q\mu\nu}^C
{\cal F}_{q\mu\nu}^C   d^4x,
\end{equation}
where

\begin{equation}
\label{eq6.14}
{\cal F}_{q\mu\nu}^C=A_{q\nu}^{-qC}
e^{\frac {A_{q\nu}^C-1} {1-q}}\partial_\mu A_{q\nu}^{C} -
A_{q\mu}^{-qC}
e^{\frac {A_{q\mu}^C-1} {1-q}}\partial_\nu A_{q\mu}^{C}+gf_{AB}^C
\sum\limits_{A,B}
e^{\frac {A_{q\mu}^A-1} {1-q}}e^{\frac {A_{q\nu}^B-1} {1-q}},
\end{equation}
leading to the motion equation

\begin{equation}
\label{eq6.15}
\sum\limits_\rho g^{\rho\rho}\partial_\rho
{\cal F}_{q\rho\sigma}^D-g\sum\limits_{\rho A C}g^{\rho\rho}
{\cal F}_{q\rho\sigma}^Cf_{AD}^C
e^{\frac {A_{q\rho}^A-1} {1-q}}=0.
\end{equation}
The field  $A_{q\mu}^C$ satisfies this equation, of course. Whenever the energy becomes low enough,
$q\rightarrow 1$, and one recovers LI and gauge invariance. .

\setcounter{equation}{0}

\section{Our  non-linear quantum Proca's q-Field}

The Proca action gives a detailed account of a massive spin-1 field of mass $m $ in a Minkowskian space-time. The associated equation is a relativistic-wave one, denominated  Proca equation. The action  is

\begin{equation}
\label{eq7.1}
{\cal S}=-\frac {1} {2}\int{\cal F}^{\dagger\mu\nu}
{\cal F}_{\mu\nu}-2m^2 A_\mu^\dagger A^\mu   d^4x,
\end{equation}
where
\begin{equation}
\label{eq7.2}
{\cal F}_{\mu\nu}=\partial_\mu A_\nu-\partial_\nu A_\mu.
\end{equation}
the motion equations being
\begin{equation}
\label{eq7.3}
(\Box+m^2)A_\mu=0\;\;\;\;\;\partial_\mu A^\mu=0.
\end{equation}

At this stage we define our q-action
\begin{equation}
\label{eq7.4}
A_{q\mu}=[1+(1-q)\ln A_\mu]^{\frac {1} {1-q}}.
\end{equation}
breaking LI. Inversion of   (\ref{eq7.4}) gives

\begin{equation}
\label{eq7.5}
A_{\mu}=e^{\frac {A_{q\mu}^{(1-q)}-1} {1-q}}
\end{equation}
From the second relation (\ref{eq7.3}) and from (\ref{eq7.5}) we find

\begin{equation}
\label{eq7.6}
\partial^\mu A_{q\mu}=0,
\end{equation}
whose associated action is

\begin{equation}
\label{eq7.7}
{\cal S}=-\frac {1} {2}\int\sum\limits_{\mu,\nu}
g^{\mu\mu}g^{\nu\nu}{\cal F}_q^{\dagger\mu\nu}
{\cal F}_{q\mu\nu}-2m^2\sum\limits_\mu
e^{\frac {A_{q\mu}^{\dagger(1-q)}-1} {1-q}}
e^{\frac {A_{q\mu}^{(1-q)}-1} {1-q}}   d^4x,
\end{equation}
with

\begin{equation}
\label{eq7.8}
{\cal F}_{q\mu\nu}=
A_{q\nu}^{-q}e^{\frac {A_{q\nu}^{(1-q)}-1} {1-q}}\partial_\mu A_{q\nu}-
A_{q\mu}^{-q}e^{\frac {A_{q\mu}^{(1-q)}-1} {1-q}}\partial_\nu A_{q\mu}.
\end{equation}
From both this and  (\ref{eq7.6})
 one finds the motion equation

\begin{equation}
\label{eq7.9}
\Box A_{q\mu}+(A_{q\mu}^{-q}-qA_{q\mu}^{-1})
\sum\limits_\nu g^{\nu\nu}(\partial_\nu A_{q\mu})^2+
m^2 A_{q\mu}^q=0,
\end{equation}
satisfied by  $A_{q\mu}$. LI is recovered in the limit  $q\rightarrow 1$.

\newpage

\setcounter{equation}{0}

\section{Conclusions}

\nd We have here obtained some new quantum  results that may be
regarded as interesting.

\nd More specifically, we have generalized to the quantum realm the classical Tsallis'
q-Schr\"{o}dinger, q-Klein-Gordon, q-Dirac, q-Proca  equations obtained in
 \cite{tp1,ts5,tz1,tz3,tz2,tp0}. We have also added equations corresponding to
q-Yang-Mills fields, Abelian and  non-Abelian.

\nd We have obtained the  q-Quantum Field Theories corresponding  to all
the above equations, and showed that in the limit  $q\rightarrow 1$ they become the customary ones.

\nd These results agree with our Nuclear Physics A previous results \cite{ts1,ts2} concerning the energies involved. One
 needs energies of up to
1 TeV in order to clearly distinguish between
q-theories and q=1, ordinary ones.

\nd All our new quantum
 q-Fields are
 q-exponential functions of the  logarithms of
the conventional $q=1$ fields. We have seen that these cannot be directly quantified because of non linearity,
 that makes not applicable the superposition principle, and then, impossible to introduce creation/annihilation operators for the q-fields. To remedy such formidable quantification obstacle we have here devised an indirect approach that has been shown to work in correct fashion.

\nd An interesting fact is that a Tsallis' q-exponential wave
 is a solution of the motion equations
(\ref{eq2.9}), (\ref{eq3.9}), (\ref{eq4.9}), (\ref{eq4.10}),
(\ref{eq5.10}), (\ref{eq6.15}), and (\ref{eq7.9}), that look quite different amongst themselves indeed!

\vskip 4mm

\nd {\bf Acknowledments} We thank Prof. A. R. Plastino for helpful discussions. We are indebted to CONICET (Argentine Agency) for economic support.

\end{document}